# Exact solution of the relationship between the eigenvalue discreteness and the behavior of eigenstates in Su-Schrieffer-Heeger lattices

Huitong Wei, [1, §] Xiumei Wang, [2, §] and Xingping Zhou [1, ‡]

[1] *Institute of Quantum Information and Technology, Nanjing University of Posts and Telecommunications, Nanjing 210003, China*

[2] *College of Electronic and Optical Engineering, Nanjing University of Posts and Telecommunications, Nanjing 210003, China*

*§ These authors contributed equally to this work.*

‡zxp@njupt.edu.cn

Eigenstate localization and bulk-boundary correspondence are fundamental phenomena in one-dimensional (1D) Su-Schrieffer-Heeger (SSH) lattices. The eigenvalues discreteness and the eigenstates localization exhibit a high degree of consistency as system information evolve. We explore the relationship between the eigenvalue discreteness and the eigenstates behavior in 1D SSH lattices. The discreteness fraction and the inverse participation ratio (IPR) combined with a Taylor expansion are utilized to describe the relationship. In the Hermitian case, we employ the bulk-edge correspondence and the perturbation theory to derive an exact solution considering both zero and non-zero modes. We also extend our analysis to the non-Hermitian cases, assuming that eigenvalues remain purely real. Our findings reveal a logarithmic relationship between the degree of eigenvalue discreteness and eigenstates localization, which holds under both the Hermitian and non-Hermitian conditions. This result is fully consistent with the theoretical predictions.

## I. INTRODUCTION

In the 1930s, renowned physicist Werner Heisenberg posed a pivotal question on the relationship between measurable quantities and intrinsic states in quantum systems in "The Physical Principles of the Quantum Theory" [1], focusing on the connection between eigenvalues and eigenstates. This framework was later expanded by Paul Dirac [2] and John von Neumann [3], laying the foundation of modern quantum mechanics and establishing a core theory of quantum measurement and system behavior. In physical systems, this theory is particularly significant, as the discreteness of eigenvalues and the behavior of eigenstates play a decisive role in determining system properties and measurement outcomes [4-8].

On the other hand, topological insulators (TIs) are materials that feature highly stable transport properties, which could lead to advancements in spintronic technologies and offer protection for quantum bits, helping prevent them from losing information due to decoherence [9]. One of the key characteristics of topological systems is the existence of localized eigenstates at the boundary between a TI and its non-topological surroundings. In the simple case of a point defect, if its eigenvalue falls deep into the band gap, then it cannot couple efficiently to the rest of the system, where no propagating mode exists at its energy. As a result, a defect eigenstates localized at this point forms, no matter whether the defect is in the band gap (zero modes) [10-17] or at the edge (non-zero modes) [18] of the system. Thus, it is obvious that there is strong mapping relationship between eigenvalues and eigenstates, whether under Hermitian or non-Hermitian conditions.

As we know, the spectrum of a periodic crystal can be characterized by the Bloch wave vector and the periodic boundary condition (PBC) is usually taken for the convenience of calculating the band structure in Su–Schrieffer–Heeger (SSH) lattices [19]. The bulk spectrum with a large enough system size is stable against boundary perturbations even though the translation invariance of the system is broken [20-23]. It explains why the bulk energy levels of a large system with open boundary conditions (OBC) can be derived from a Bloch band calculation, namely the PBC condition. Although it is challenged in the non-Hermitian systems, the non-Bloch band theory [24, 25], biorthogonal bulk-boundary correspondence [26], and the non-Bloch Band theory in arbitrary dimensions [27] are proposed to solve this problem. These bulk–boundary correspondence methods will help us to catch the relationship between the eigenvalue discreteness and the behavior of eigenstates.

Topological states also have attracted significant attention in photonic devices due to their novel approach in molding the flow of light. Topology-driven localization of optical states unlocks special prospects to facilitate lasing [28-32] and quantum light generation [33, 34] in photonic structures with improved reliability. Among those pioneering studies, one-dimensional (1D) SSH designs are employed to demonstrate lasing generation driven by topological states [30-32]. It can be seen that the behavior of topological eigenstates plays a key role in the lasing process.

In this work, we systematically analyze the relationship between the eigenvalue discreteness and the eigenstates behavior in the topologically non-trivial phase of the 1D SSH model, and validate the findings in both Hermitian and non-Hermitian cases. The model incorporates an odd-numbered node configuration with a central defect, introducing topological features and complex mode distributions, such as zero modes and non-zero modes. We define the discreteness fraction $D$, and utilize a Taylor expansion to relate the inverse participation ratio ($IPR$) [35, 36] to the model's structural information, thereby quantifying the eigenvalue discreteness and eigenstates localization. In the Hermitian case, we approximate the discreteness fraction by utilizing the bulk-edge correspondence, which links the discreteness to structural information. Notably, in the analysis of non-zero modes, we employ the perturbation theory to investigate their behavior and characteristics. Then we extend this analysis to the non-Hermitian system, focusing on the case where eigenvalues remain purely real despite the introduction of gain-loss mechanism. Our results demonstrate that the degree of eigenvalue discreteness and eigenstates localization exhibit a logarithmic relationship under both Hermitian and non-Hermitian conditions. The analytical solutions are fully consistent with the theoretical predictions, providing robust validation of the inherent connection between eigenvalue discreteness and eigenstates localization.

The paper is structured as follows: In section II, we introduce the model and methods for analyzing the numerical relationship between the eigenvalue discreteness and eigenstates localization. In Sec. II.1, we discuss the 1D Hermitian and non-Hermitian SSH models, and in Sec. II.2, we present two indicators for quantifying discreteness and localization. In Sec III, we systematically analyze the classification and relationship of discrete eigenvalue points in the model. Specifically, in Sec III.1 and Sec III.2 we explore the properties of these points under Hermitian and non-Hermitian conditions, respectively. In Sec III.1&2.A, we derive the relationships for zero modes, while in Sec III.1&2.B, we focus on non-zero modes. Finally, in Sec. III.1&2 C we validate the theoretical findings with analytical solutions, confirming the reliability of the results.

# II. MODEL AND METHOD

## 2.1Model

We first consider the SSH model shown in Fig. 1(a). Its Hamiltonian is expressed as:

$$H = \sum_l [t_1 c_l^\dagger c_l + t_2 c_{l+1}^\dagger c_l + i\gamma (c_l^\dagger c_l - c_l^\dagger c_l)] \qquad (1)$$

Here, $c_l^\dagger$ and $c_l$ represent the creation and annihilation operators at the $l$-th lattice respectively, $t_1$ (intracell) and $t_2$ (intercell) indicate the and hopping strengths respectively, $i\gamma$ describes the gain and loss mechanism at the lattice sites. The hopping strengths are defined as $t_1 = 1 + \alpha\cos\theta$ and $t_2 = 1 - \alpha\cos\theta$, where $\alpha$ is the modulation parameter (set to $0.5$), and $\theta$ is the modulation phase in the range of $[0, 2\pi)$.

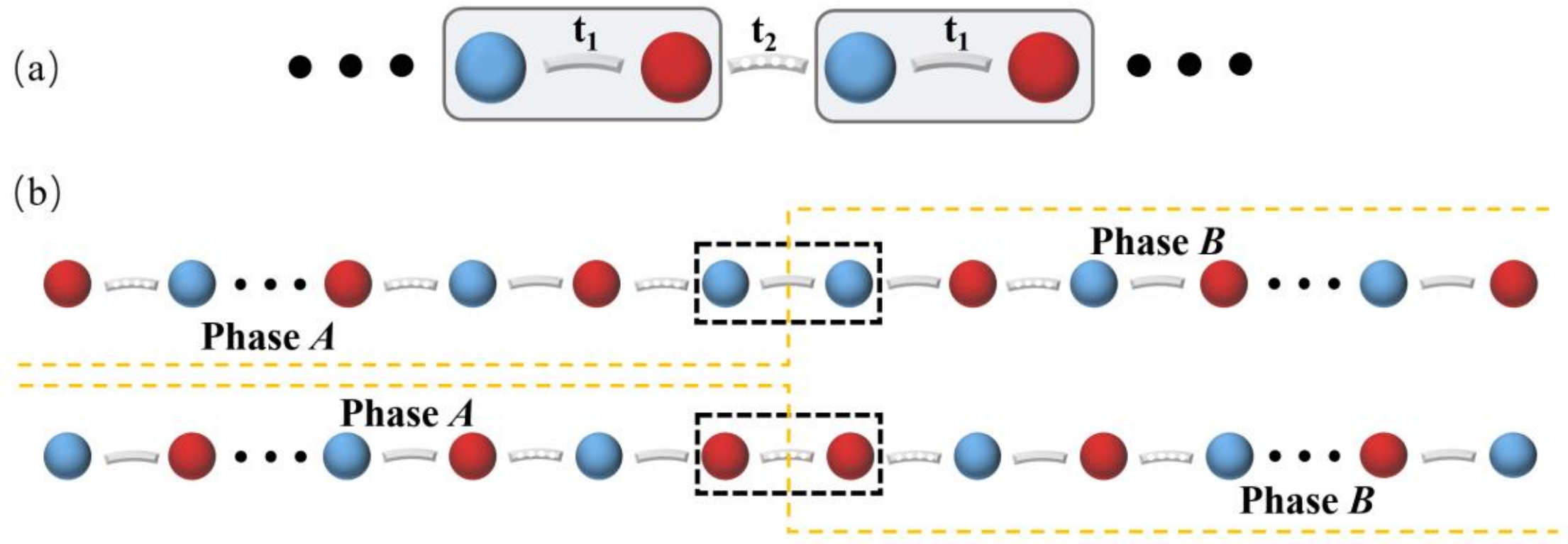

FIG. 1. (a) The structure of the 1D SSH model: The red and blue circles represent two sublattice sites, while the grey links indicate the hopping strengths $t_1$ and $t_2$. The arrows with tildes illustrate the gain-loss effects. (b) Topological phases and boundary defects: The black dashed boxes mark the kink defects at the boundaries between Phase *A* and Phase *B*.

When $\pm i\gamma = 0$, the model is the standard Hermitian model, which describes a 1-D alternating hopping structure. This model can exhibit rich topological features under different settings [37, 38]. As shown in Fig. 1(b), different hopping strengths correspond to the topological phase regions of the model: Phase *A* and Phase *B* represent different relative strength of hopping, forming two different types of topological phases. The structure of the topological phases determines the topological properties of the system and influences the existence and distribution of boundary states. The yellow dashed boxes in Fig. 1(b) illustrate the distribution of topological boundary states across different phases.

When $\pm i\gamma \neq 0$, this model evolves into a non-Hermitian SSH model. Gain and loss mechanisms break the Hermiticity of the model, extending the eigenvalue spectrum from the real domain to the complex domain. Similar to the Hermitian SSH model, this model is still constructed with alternating strong bonds $t_1$ and weak bonds $t_2$, but the introduction of non-Hermitian terms significantly alters the topological properties of the system. The interfaces between Phase *A* and Phase *B* exhibit a more complex distribution of boundary states, with the non-Hermitian skin effect becoming a key characteristic of the system [17, 39-42].

## 2.2Methods

In this section, we introduce two essential indicators to characterize the physical properties of the model: the degree of eigenstates localization and eigenvalue discreteness. The localization is measured by the *IPR*, which reflects the spatial localization behavior of the eigenstates. To quantify the uniformity and discreteness of eigenvalue within a neighborhood lattice, we define the discreteness fraction *D*, which describes their distribution characteristics. Through a comprehensive analysis of these two indicators, we systematically investigate the relationship between the discreteness of eigenvalues and the localization of eigenstates. A detailed exploration of both indicators and their respective approximations is presented, elucidating their relationship and offering valuable insights into the physical properties of the SSH model.

### A. Behavior of Localization

For the Hamiltonian in the Hermitian SSH model described in Eq. (1), the eigenvalue equation is given by $H\psi = E\psi$ , where $\psi = [\psi_1, \psi_2, ..., \psi_N]^T$ . The eigenvalue equation consists of the following series of equations [23, 42], including the bulk equations:

$$t_1\psi_{i-1} - E_i\psi_i + t_2\psi_{i+1} = 0, i = 2,...., L-1 \quad (2)$$

and the boundary equations:

$$t_1\psi_2 - E\psi_1 = 0 \quad (3)$$

$$t_1\psi_{L-1} - E\psi_L = 0$$

Assuming the eigenstates $\psi_i$ has the exponential form $\psi_i = \lambda^i$ and substituting it into Eq. (2), the eigenvalue equation can be obtained as:

$$t_1\lambda^{-1} - E + t_2\lambda = 0 \quad (4)$$

From this equation, the expression for $\lambda$ can be derived as:

$$\lambda = \frac{E \pm \sqrt{E^2 - 4t_1t_2}}{2t_2} \tag{5}$$

Generally, the condition $|\lambda|<1$ is adopted to ensure the physical validity of the eigenstates distribution in the analysis of localized states, as it corresponds to exponential decay [25]. Our analysis primarily focuses on the spatial distribution of the eigenstates, which remains independent of the magnitude of the eigenvalue *E*. Therefore, the approximate magnitude of $\lambda$ under the parameter relationship $|\lambda|<1$ can be expressed as:

$$|\lambda| \sim \sqrt{\frac{t_1}{t_2}} \tag{6}$$

Consequently, the spatial distribution of the eigenstates at the $i$-th site is given by:

$$\psi_i = |\lambda|^i \sim (\sqrt{\frac{t_1}{t_2}})^i \tag{7}$$

This expression clearly indicates that the localization behavior of the eigenstates is determined by the hopping strength ratio $t_1/t_2$. When $|\lambda|<1$, the model presents topological edge states, and the spatial distribution follows an exponential decay.

To quantify the degree of localization, we introduce the *IPR*, an important indicator used to characterize the spatial localization of states [35, 36]. Based on Eq. (7), the expression for *IPR* is given as:

$$IPR = \sum_i |\psi_i|^4 \sim \sum_i (-r)^{2i} \tag{8}$$

where $r = -t_1/t_2$ represents the hopping strengths ratio. For a finite lattice chain with sufficiently large *L*, the *IPR* can be approximated as:

$$IPR \sim \frac{1-r^{2(L+1)}}{1-r^2} \approx \frac{1-r^{2L}}{1-r^2} \tag{9}$$

By rearranging Eq. (9), we can solve for $r$:

$$r \sim [1 - IPR(1-r^2)]^{1/(2L)} \tag{10}$$

When $IPR \to 0$, it corresponds to $r \to 0$, indicating delocalization of the eigenstates. As the *IPR* increases, from Eq. (9), we find:

$$r \to (1 - IPR)^{1/(2L)} \tag{11}$$

According to the Taylor expansion of the exponential function $e^x = 1 + x + x^2/2! + x^3/3! + ...$, we approximate $e^{-IPR}$ as:

$$e^{-IPR} \approx 1 - IPR + \frac{IPR^2}{2!} - \frac{IPR^3}{3!} + ... \quad (12)$$

For 1-*IPR*, the expression can be obtained as:

$$1 - IPR \approx e^{-IPR} + \frac{IPR^2}{2!} - \frac{IPR^3}{3!} + ... \quad (13)$$

Due to the definition, the *IPR* lies within the range of (0,1), which implies $1 - IPR \in (0,1)$. When *IPR* is approaching to 0, the Taylor expansion of $e^{-IPR}$ converges rapidly around the expansion point [43], and higher-order terms (e.g., $IPR^2/2!$) can be negligible. Thus, the approximation can be further simplified to:

$$1 - IPR \approx e^{-IPR} \quad (14)$$

Substituting this into Eq. (11), we obtain the final approximate expression when $r$ is not close to 0:

$$IPR = -2L\ln r \quad (15)$$

The above approximation indicates that the degree of localization can be directly estimated by the exponential decay of the eigenstates. This estimation is valid under conditions such as large chain length *L* and relatively large *IPR*.

For the eigenstates in the non-Hermitian SSH model, we consider the following bulk equation [44]:

$$t_1\psi_{i-1} - (E_i - i\gamma)\psi_i + t_2\psi_{i+1} = 0, i = 2, ...., L-1 \quad (16)$$

where $-i\gamma$ is the imaginary term introduced by the non-Hermitian gain-loss mechanism. Assuming the eigenstates takes the form $\psi_i = \lambda^i$ and substituting this into Eq. (16), the eigenvalue equation can be obtained as:

$$t_1\lambda^{-1} - (E - i\gamma) + t_2\lambda = 0 \quad (17)$$

Solving this equation yields the expression for $\lambda$:

$$\lambda = \frac{E - i\gamma \pm \sqrt{(E - i\gamma)^2 - 4t_1t_2}}{2t_2} \quad (18)$$

In this case, the introduction of the imaginary term $i\gamma$ affects only the magnitude of the eigenvalue, without altering the exponential decay behavior of the eigenstates. Therefore, the spatial distribution of the eigenstates in the non-Hermitian SSH model remains identical to the Hermitian case as long as the model retains *PT*-symmetry. By directly analyzing this process, it can be shown that the results are consistent with those in Eq. (6) for the Hermitian case. Consequently, through the same derivation, the *IPR* in the non-Hermitian system can also be exactly expressed as shown in Eq. (15).

### B. Eigenvalue Discreteness

We introduce the discreteness fraction *D* as a new defined indicator. Discrete eigenvalue points are identified as those whose differences exceed a certain threshold value. We set the definition of *D* as follows:

$$D = \sum_{layer=1}^{n_{layer}} (|E(i) - E(i-layer)| + |E(i+layer) - E(i)|) \tag{19}$$

Here， $E(i)$ represents the $i$-th eigenvalue， and $n_{layer}$ denotes the number of layers in the neighborhood. This definition calculates the absolute differences of eigenvalues $E(i)$ within the neighborhood lattice, effectively quantifying their discreteness. To ensure that the calculated discreteness differences are significant enough to represent the main contribution to the fraction, we further define the threshold *thr* as：

$$thr = \frac{\beta \times \max(diff_k)}{neighborhood}, \quad diff_k = |E_{k_1} - E_{k_2}| \tag{20}$$

where $diff_k$ represents the difference between any two eigenvalues within the neighborhood, $\beta$ is an adjustment parameter, and *neighborhood* denotes the range of the neighborhood. Only those eigenvalue differences that satisfy the condition $|E(i) - E(i-layer)| > thr$ are included in the calculation of *D*. Finally, the discreteness fraction *D* of eigenvalues can be approximately expressed as:

$$D \sim \Delta E_k = E_{k_1} - E_{k_2} \tag{21}$$

Here, $\Delta E_k$ represents eigenvalue differences. $E_{k_1}$ and $E_{k_2}$ are the two eigenvalues in the neighborhood whose difference exceeds *thr*. The magnitude of the *D* directly reflects the degree of discreteness in the eigenvalue distribution: when the *D* is large, the eigenvalues exhibit a high level of discreteness; in contrast, when the *D* is small, the eigenvalue distribution is relatively uniform.

## III. RELATIONAL DERIVATION

In our model, the boundary conditions and the topological phase boundaries formed at the junctions of Phase *A* and Phase *B* collectively influence the distribution of eigenvalues. This results in two primary types of discrete eigenvalue points: zero modes and non-zero modes [12, 37]. Here, the lattice size of 79 is chosen to ensure clear topological boundary phenomena, with its size separating boundary and bulk states, and its oddness guaranteeing zero modes.

As shown in Fig. 2(a), discrete eigenvalue points induced by boundary conditions manifest as zero modes (e.g., Mode 40), where the eigenvalues are located near the center of the bandgap. These points reflect the localized behavior of the system at the boundaries. In contrast, discrete eigenvalue points arising from topological phase boundaries vary with parameter changes. When $t_1 > t_2$, i.e., $\theta \in [0, \frac{\pi}{2}) \bigcup [\frac{3\pi}{2}, 2\pi)$, the discrete eigenvalues are located at the center of the bandgap, expressing as zero modes (e.g., Modes 39, 40, and 41). Conversely, when $t_1 < t_2$, i.e., $\theta \in [\frac{\pi}{2}, \frac{3\pi}{2})$, these eigenvalues shift to the edges of the energy band, expressing as non-zero modes (e.g., Modes 1 and 79), which are the discrete eigenvalue points distinct from the 'zero mode'. Next, we will conduct an in-depth analysis of the physical properties of these two types of discrete eigenvalue points.

## 3.1.Hermitian Cases

### A. Zero Modes

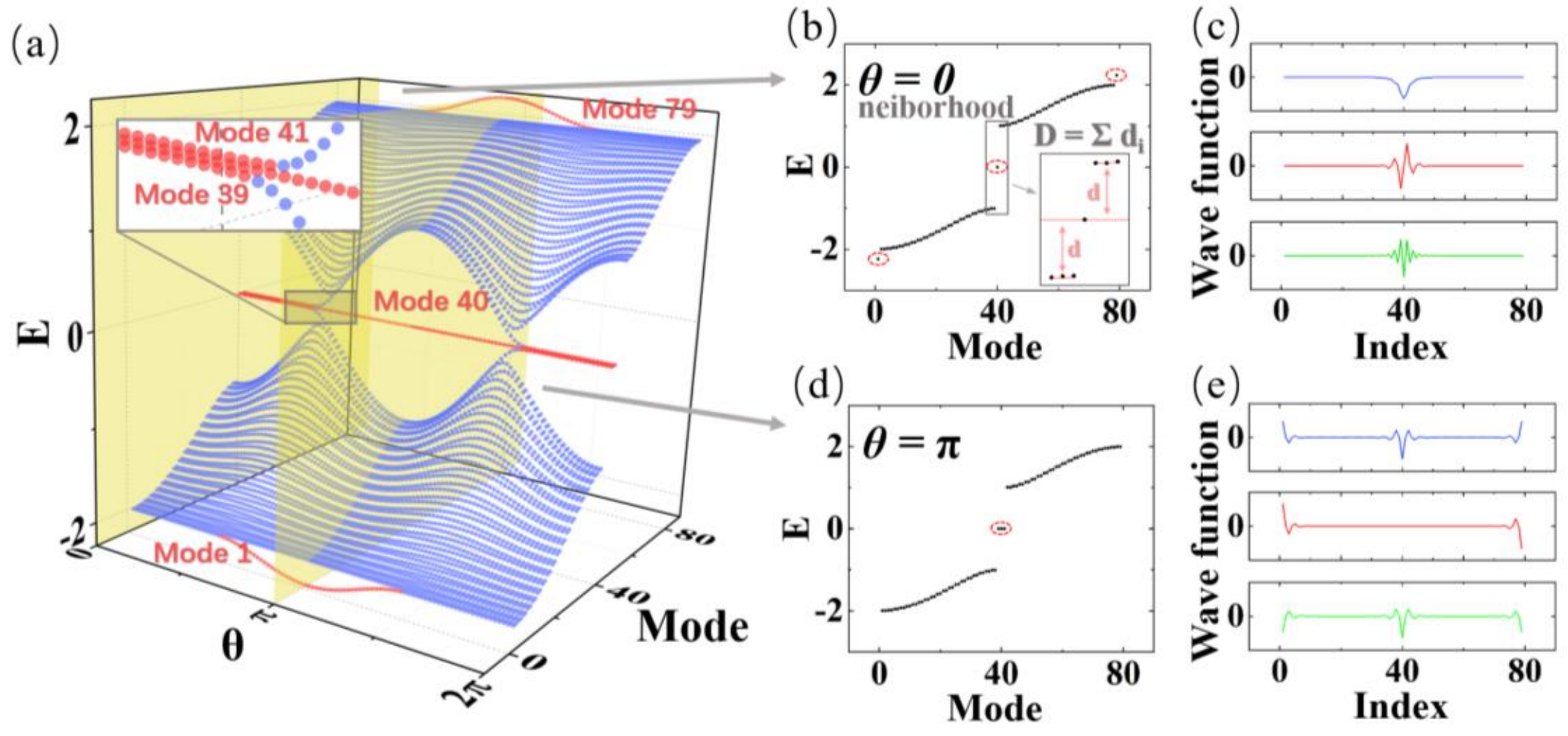

FIG. 2. (a) Energy spectrum as a function of θ, with red markers highlighting discrete eigenvalues (Modes 1, 39, 40, 41, and 79). The inset zooms in on Modes 39 and 41 near the band edge, illustrating their discreteness. (b) Eigenvalue at $\theta=0$, with the inset showing the calculation of the discreteness fraction $D$. (c) Eigenstates of Modes 1, 40, and 79 at $\theta=0$, demonstrating localization and delocalization. (d) Eigenvalue at $\theta=\pi$, where edge eigenvalues persist. (e) Eigenstates of Modes 39, 40, and 41 at $\theta=\pi$, further confirming their localization.

Under PBC, the eigenvalues of the SSH model can be expressed in terms of the momentum *k* as:

$$E_k = \pm\sqrt{t_1^{\ 2} + t_2^{\ 2} + 2t_1 t_2 \cos k} \tag{22}$$

Then, the difference of these eigenvalues can be simplified as:

$$\Delta E_k = E_{k_1} - E_{k_2} = E_{k_1} \mp \sqrt{t_1^{\ 2} + t_2^{\ 2} + 2t_1 t_2 \cos k_2} \tag{23}$$

From Eq. (23), the discreteness fraction D in Eq. (21) can be expressed as a function of the hopping strength $t_1$, $t_2$, and the momentum k.

Under OBC, the eigenvalues corresponding to localized states are predominantly influenced by boundary effects, rather than being determined by the band energy spectrum of the PBC model in the Hermitian case. Consequently, as shown in Fig. 2(b) and Fig. 2(d), these zero modes manifest the model's topological properties, which are further illustrated in Fig. 2(c) and Fig. 2(e). They exhibit the following distribution characteristics: (1) The zero-mode eigenvalues $E_k$ are typically close to 0; (2) The eigenvalue points corresponding to localized states typically appear near the bandgap edges, where the discreteness of eigenvalues is significantly enhanced. Specifically, the index layers of eigenvalues which exceed the threshold value *thr* are generally distributed near the lower edge of the bandgap. Based on these characteristics, we derive the following relationship:

$$E_{k_1} \approx 0, \quad k_2 \approx \pm\pi \tag{24}$$

For the zero modes, the eigenvalue difference can be further simplified to:

$$\Delta E_k \approx 0 \mp \sqrt{t_1^{\ 2} + t_2^{\ 2} - 2t_1 t_2} = | t_1 - t_2 | \tag{25}$$

Thus, by combining with Eq. (21), the expression for the discreteness fraction *D* can be written as:

$$D \sim t_2 \, | r - 1 | \tag{26}$$

Here, $r$ is the coupling coefficient ratio defined as $t_1 / t_2$. Substituting it into the approximated expression for *IPR* in Eq. (15), we obtain:

$$D \sim t_2 (e^{-IPR/(2L)} - 1) \tag{27}$$

The final relationship between the *IPR* and *D* is derived after further simplification.:

$$IPR \sim -2L \ln(\frac{D}{t_2} + 1) \sim \ln D \tag{28}$$

It indicates that $D$ and $IPR$ exhibit a nearly linear logarithmic relationship, when both of them are relatively large.

## B. Non-Zero Modes

We introduce the perturbation theory [45] to systematically analyze the distribution of eigenvalue discreteness in non-zero modes and their relationship with the structure information. Based on the first-order perturbation analysis, we conduct a quantitative analysis of eigenvalue distribution near the model's boundary. Specifically, we define $H_0$ as the unperturbed Hamiltonian matrix. When a perturbation is introduced into the model, the Hamiltonian is modified as:

$$H = H_0 + \Delta H \tag{29}$$

$$H = \begin{pmatrix} \ddots & \ddots & 0 & 0 & 0 & 0 & 0 & 0 & \cdots \\ \ddots & 0 & t_1 & 0 & 0 & 0 & 0 & 0 & 0 \\ 0 & t_1 & 0 & t_2 & 0 & 0 & 0 & 0 & 0 \\ 0 & 0 & t_2 & 0 & t_1 & 0 & 0 & 0 & 0 \\ 0 & 0 & 0 & t_1 & 0 & t_1 & 0 & 0 & 0 \\ 0 & 0 & 0 & 0 & t_1 & 0 & t_2 & 0 & 0 \\ 0 & 0 & 0 & 0 & 0 & t_2 & 0 & t_1 & 0 \\ 0 & 0 & 0 & 0 & 0 & 0 & t_1 & 0 & \ddots \\ \cdots & 0 & 0 & 0 & 0 & 0 & 0 & \ddots & \ddots \end{pmatrix}$$

$$H_0 = \begin{pmatrix} \ddots & \ddots & 0 & 0 & 0 & 0 & 0 & 0 & \cdots \\ \ddots & 0 & t_1 & 0 & 0 & 0 & 0 & 0 & 0 \\ 0 & t_1 & 0 & t_2 & 0 & 0 & 0 & 0 & 0 \\ 0 & 0 & t_2 & 0 & t_1 & 0 & 0 & 0 & 0 \\ 0 & 0 & 0 & t_1 & 0 & t_2 & 0 & 0 & 0 \\ 0 & 0 & 0 & 0 & t_2 & 0 & t_1 & 0 & 0 \\ 0 & 0 & 0 & 0 & 0 & t_1 & 0 & t_2 & 0 \\ 0 & 0 & 0 & 0 & 0 & 0 & t_2 & 0 & \ddots \\ \cdots & 0 & 0 & 0 & 0 & 0 & 0 & \ddots & \ddots \end{pmatrix}$$

$$\Delta H = \begin{pmatrix} \ddots & \ddots & 0 & 0 & 0 & 0 & 0 & 0 & \cdots \\ \ddots & 0 & 0 & 0 & 0 & 0 & 0 & 0 & 0 \\ 0 & 0 & 0 & 0 & 0 & 0 & 0 & 0 & 0 \\ 0 & 0 & 0 & 0 & 0 & 0 & 0 & 0 & 0 \\ 0 & 0 & 0 & 0 & 0 & t_1 - t_2 & 0 & 0 & 0 \\ 0 & 0 & 0 & 0 & t_1 - t_2 & 0 & t_2 - t_1 & 0 & 0 \\ 0 & 0 & 0 & 0 & 0 & t_2 - t_1 & 0 & t_1 - t_2 & 0 \\ 0 & 0 & 0 & 0 & 0 & 0 & t_1 - t_2 & 0 & \ddots \\ \cdots & 0 & 0 & 0 & 0 & 0 & 0 & \ddots & \ddots \end{pmatrix}$$

Here, $\Delta H$ represents the perturbation matrix induced by the defect and its expression is given by:

$$\Delta H = (t_1 - t_2)\sum_{\mathrm{i}=L/2}^{L-1} (|i\rangle\langle i+1| + |i+1\rangle\langle i|) \qquad (30)$$

Based on the perturbation theory, we assume that the $H_0$ has an eigenvalue $E_k^0$, with its eigenvector denoted as $\psi_i^0$. The perturbed eigenvalue is expressed as:

$$E_k = E_k^0 + \Delta E_k^0 \qquad (31)$$

The difference of the eigenvalue $\Delta E_k^0$ is given by:

$$\Delta E_k^0 = \langle \psi_i^0 \,|\, \Delta H \,|\, \psi_j^0 \rangle = |t_1 - t_2| \cdot \sum_{i=L/2}^{L-1} \psi_i^0 \psi_{i+1}^0 \qquad (32)$$

where the overlap integral of the eigenstates $\sum_{i=L/2}^{L-1} \psi_i \psi_{i+1}$ is a constant and denoted as $A$ under the condition of a fixed eigenvalue $E_k^0$. Consequently, the difference of the eigenvalue can be simplified as:

$$\Delta E_k^0 = A \cdot |t_1 - t_2| \qquad (33)$$

For the non-zero modes, the approximate expression for the eigenvalue difference $\Delta E_k$ is given by:

$$\Delta E_k = E_{k_1} - E_{k_2} = E_{k_L} - E_{k_{L-1}} \qquad (34)$$

Here, $E_{k_L}$ represents the eigenvalue at the edge of the energy band, and $E_{k_{L-1}}$ corresponds to the eigenvalue adjacent to the non-zero mode.

Under OBC, the eigenvalues exceeding the threshold index layer are usually distributed near the upper and lower boundaries of the band edges, as indicated by Mode 1 and 79 in Fig.2 (b). The corresponding momentum satisfies:

$$k_L^0 \approx 0 \text{ or } k_L^0 \approx 2\pi \qquad (35)$$

When the perturbation is introduced, the edge eigenvalue points exhibit enhanced separation. At this time, $k_{L-1}$ is replaced by $k_L^0$, which becomes the momentum at the band edge. It leads to the following relationships:

$$k_L^0 = k_{L-1},\ E_{k_L}^0 = E_{k_{L-1}} \qquad (36)$$

Here, $k_L^0$ represents the momentum of the non-zero modes before the perturbation, while $k_{L-1}$ denotes the momentum of the eigenvalue adjacent to the non-zero mode after the perturbation. Thus, combining Eqs. (34-36), the eigenvalue difference can be further simplified as:

$$\Delta E_k = E_{k_L}^0 + \Delta E_{k_L}^0 - E_{k_{L-1}} = \Delta E_{k_L}^0 = A \cdot |t_1 - t_2| \tag{37}$$

Combining Eq. (21), the approximate expression for the discreteness fraction $D$ can be further written as:

$$D \sim A \cdot t_2 |r-1| \tag{38}$$

Then, it can be observed that the expression for the discreteness fraction $D$ of the non-zero modes is consistent with the counterpart of zero modes. By employing a similar derivation, the relationship between *IPR* and $D$ ultimately is equal to Eq. (28) for zero modes.

**C. Analytical Solutions**

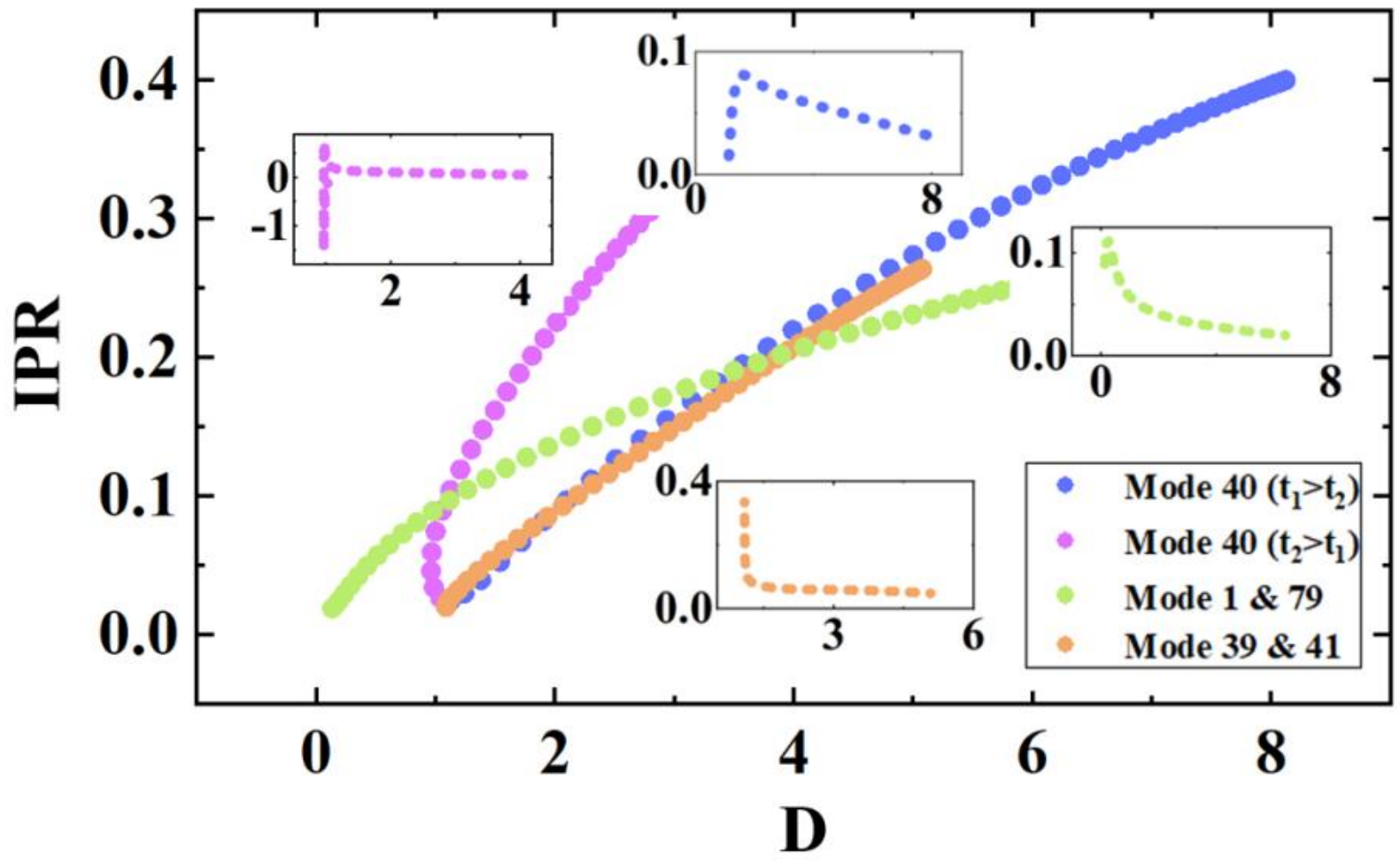

FIG. 3. The relationship between *IPR* and *D* for different modes in the Hermitian case. Insets show the reciprocals of the corresponding colored curves

Through the theoretical derivation of the relationship between $D$ and *IPR*, we reveal the degree of eigenvalues discreteness (as shown in Fig. 2(b) and (d)) exhibits a logarithmic relationship with the degree of eigenstates localization (as shown in Fig. 2(c) and (e)) under the Hermitian case. This conclusion is further verified by the numerical results presented in Fig. 3 Specifically, in the functional relationship between *IPR* and $D$, it is clearly observed that when $D$ is relatively small, the system resides in a delocalized state, where the relationship between *IPR* and $D$ feels like random scattering. As the discreteness fraction increases, $D$ and *IPR* exhibit an increasingly obvious logarithmic relationship, expressed as *IPR*~ln$D$. Furthermore, the inset of Fig.

3 revels a characteristic inverse dependence 1/x, highlighting the tight coupling between the eigenvalue discreteness and the eigenstate localization.

## 3.2. Non-Hermitian Cases

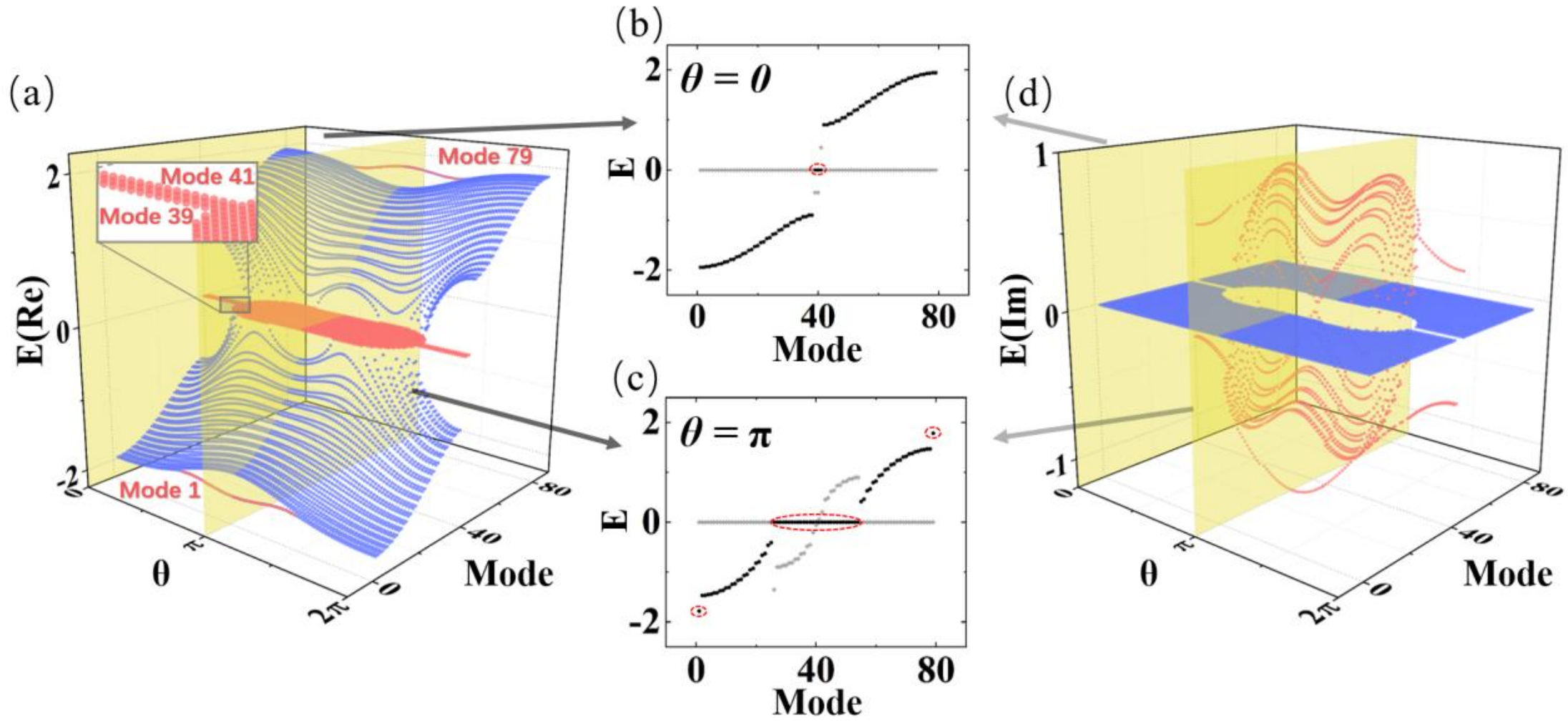

FIG. 4. (a) Overview of the real part of the energy spectrum, with red markers indicating non-zero modes. (b), (c) Detailed mode behavior at $\theta=0$ and $\theta=\pi$, respectively. (d) Evolution of the imaginary energy spectrum, focusing on regions associated with PT-symmetry breaking.

In the non-Hermitian SSH model, the real and imaginary components of the eigenvalues characterize the model's *PT*-symmetric behavior in different parameter regimes. The eigenvalues $E_k$ under PBC is determined by the gain-loss parameter and the hopping strengths of the model, as described by:

$$E_k = \pm\sqrt{t_1^{\ 2} + t_2^{\ 2} + 2t_1 t_2 \cos k - (\frac{\gamma}{2})^2} \qquad (39)$$

where $\gamma$ denotes the gain-loss strength, which plays a critical role in shaping the distribution of the eigenvalues.

As shown in Fig. 4(a) and (d), when $\gamma / 2$ is relatively small, the model maintains *PT*-symmetry despite the introduction of gain and loss. In this regime, the imaginary part of the eigenvalues is zero and the model retains its topological protection of energy bands. However, as $\gamma / 2$ exceeds a critical value—the eigenvalue equation no longer satisfies the reality condition—*PT*-symmetry is spontaneously broken, and the eigenvalues begin to acquire imaginary components. We focus on the case where *PT*-symmetry remains unbroken, and the eigenvalues are purely real. Under these conditions, the critical relationship between the gain-loss parameter and the hopping strengths of the system is given by:

$$t_1^2 + t_2^2 + 2t_1t_2\cos k - (\frac{\gamma}{2})^2 > 0 \tag{40}$$

**A. Zero Modes**

For the zero modes, the eigenvalues approach 0 ( $E_{k_1} \approx 0$ ) while the discontinuity in the distribution of eigenvalues near the band edge ( $k_2 \approx \pm\pi$ ) becomes more pronounced under the condition specified by Eq. (40). Referring to Eq. (39), the eigenvalue difference under OBC can be simplified as:

$$\Delta E_k = E_{k_1} - E_{k_2} = 0 \mp \sqrt{t_1^2 + t_2^2 - 2t_1t_2 - (\frac{\gamma}{2})^2} = \sqrt{t_2^2(r-1)^2 - (\frac{\gamma}{2})^2} \tag{41}$$

Consequently, by combining Eq. (21), the final expression for the discreteness fraction *D* is given by:

$$D \sim \sqrt{t_2^2(r-1)^2 - (\frac{\gamma}{2})^2} \tag{42}$$

By substituting the approximate expression of *IPR* in Eq. (15), the relationship between the discreteness fraction *D* and *IPR* can be expressed as:

$$D \sim \sqrt{t_2^2(e^{-\frac{IPR}{2L}} - 1)^2 - (\frac{\gamma}{2})^2} \tag{43}$$

Further simplification yields the final approximate relationship between *IPR* and *D* as:

$$IPR \sim -4L\ln(1 + \sqrt{\frac{\gamma^2}{4t_2^2} + \frac{D^2}{t_2^2}}) \sim \ln D \tag{44}$$

This equation indicates that when both *D* and *IPR* are relatively large, while, $t_1$, $t_2$ and $\gamma/2$ satisfy the conditions in Eq. (40), there exists an approximate logarithmic relationship between *D* and *IPR*.

**B. Non-Zero Modes**

In the non-Hermitian case, the degree of the eigenvalues discreteness at the band edges can similarly be quantitatively described by the perturbation theory. Similar to the Hermitian case, the eigenvalue difference can be expressed as Eq. (37). Consequently, the expression for the discreteness fraction *D* of non-zero modes in the non-Hermitian case remains consistent with the results in the Hermitian case. By following a similar derivation, the final approximate relationship between *IPR* and *D* is also consistent with the Hermitian case, as given by Eq. (28).

This result indicates that even under non-Hermitian conditions, the introduction of the gain-loss parameter $\gamma$ does not alter the fundamental law between the eigenvalue discreteness and the eigenstates localization.

### C. Analytical Solutions

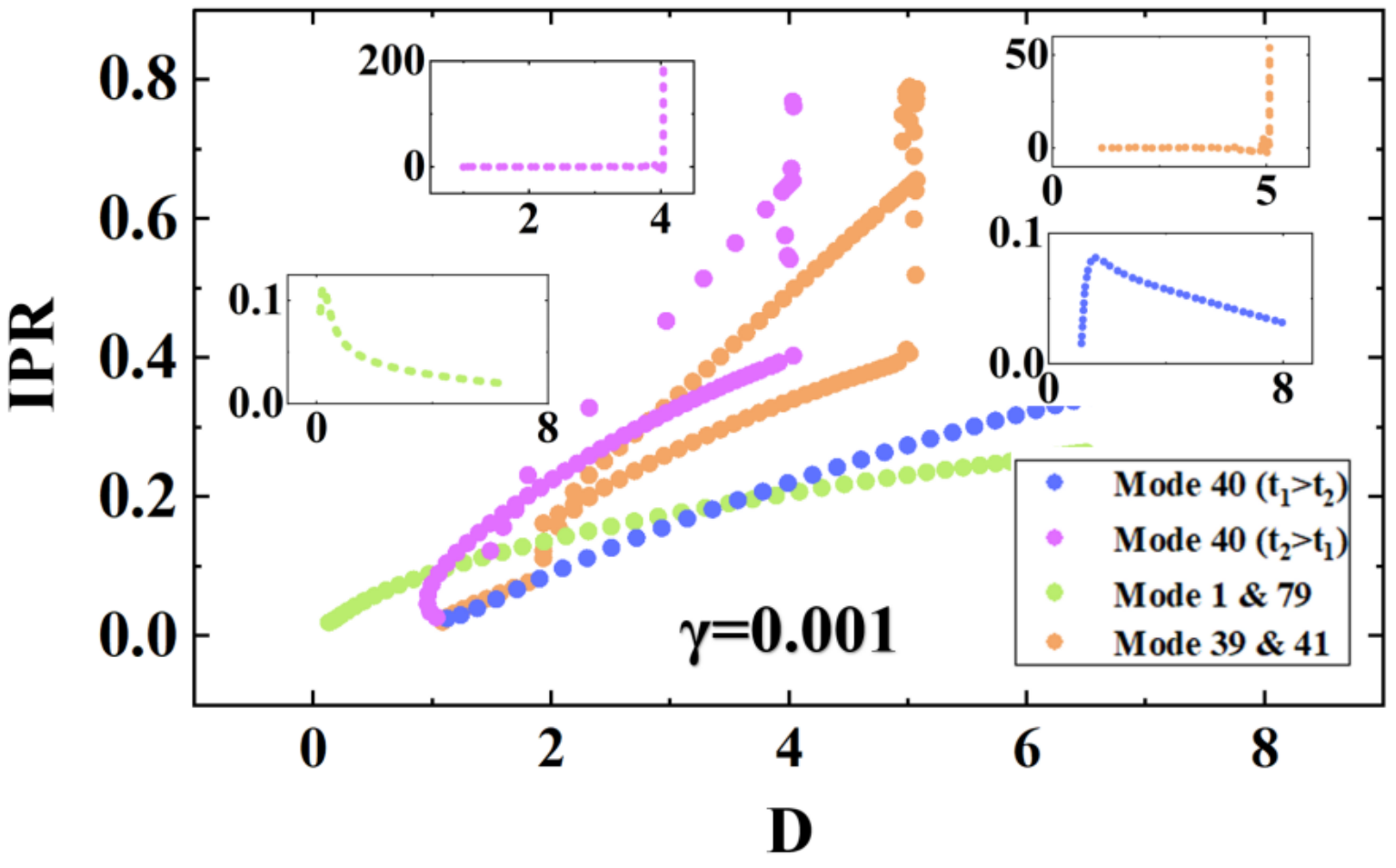

FIG. 5. The relationship between *IPR* and *D* for different modes in the non-Hermitian case. Insets show the reciprocals of the corresponding colored curves

Through the theoretical derivation, we reveal that when $t_1$, $t_2$ and $\gamma/2$ satisfy the conditions of Eq. (40), *D* and *IPR* exhibit a logarithmic relationship in the non-Hermitian case. This conclusion is further validated by the numerical results presented in Fig. 5. For the zero modes (Mode 39, 40 and 41 in Fig. 4(a)), the relationship between *IPR* and *D* generally follows a logarithmic dependence, demonstrating a stable relationship between eigenvalue discreteness and eigenstates localization under certain conditions. While the coupling parameter varies, some eigenvalues deviate from this relationship, forming the so-called “runaway” phenomenon. It occurs when $t_1$, $t_2$ and $\gamma/2$ fails to satisfy the conditions of Eq. (40). At this time, the non-Hermitian effect becomes prominent, leading to abnormal behavior in the eigenvalue distribution.

In contrast, for the paired non-zero modes (Mode 1 and 79 in Fig. 4(a)), the relationship between *IPR* and *D* remains consistent with the Hermitian case, following a strict logarithmic dependence. This behavior persists as long as gain-loss strength is not too large to break topological protection. It suggests that the non-zero modes exhibit stronger localization and stability under variations in coupling and gain-loss parameters.

As a result, the model retains its topological protection within a well-defined parameter range, ensuring robustness against perturbations in the gain-loss strength.

## IV. CONCLUSIONS

In conclusion, we have proposed the discreteness fraction *D* of the eigenvalues as a newly defined indicator and explored its relationship with the behavior of eigenstates in 1D SSH lattices. With the help of a Taylor expansion, we establish the relationship with the *IPR* successfully. Then, the bulk-edge correspondence is applied to link the discreteness of zero modes with the structural information in the Hermitian condition. For the non-zero modes, we employ the perturbation theory to conduct a quantitative analysis of eigenvalue distribution. Up to now, the exact solution of the relationship between the eigenvalue discreteness and the behavior of eigenstates in 1D SSH lattices has been established. We further extend it to the non-Hermitian case. The exact solution matches well with the theoretical prediction, in which the degree of eigenvalue discreteness and eigenstates localization exhibit a logarithmic relationship under both the Hermitian and non-Hermitian conditions. We hope the results can help to create and manipulate the topological states in photonics [46, 47] and beyond [48-50].

Our theoretical framework can be extended to the two-dimensional (2D) SSH model. In 2D models, we hypothesize that there may still be a mathematical connection between the discreteness fraction of the eigenvalues and eigenstates behavior. However, due to the more intricate topological phenomena (corner states and edge states) in 2D systems [51-53] or higher dimensions [54, 55], the behaviors of eigenvalue discreteness and eigenstate localization are likely to become more complex and diverse compared to their 1D counterparts. Thus, further works must be relied on more precise mathematical and physical methods to uncover the detailed exact solution.

**Acknowledgements**

The authors thank for the support by National Natural Science Foundation of China under (Grant 12404365).